\def\hcorrection#1{\advance\hoffset by #1 }
\def\vcorrection#1{\advance\voffset by #1 }

\documentstyle[11pt]{article}

\vcorrection{-0.70in}
\hcorrection{-0.60in}
\begin{document}

\title{On pressure and velocity flow boundary  conditions 
and 
bounceback for the lattice Boltzmann BGK model}
\author{
Qisu Zou \thanks{%
 Theoretical Division, Los Alamos National Lab, Los Alamos, NM 87545} 
\thanks{%
 Dept. of Math., Kansas State University, Manhattan, KS 66506} 
and Xiaoyi He \thanks{%
Center for Nonlinear Studies, Los Alamos National Lab } \thanks{%
Theoretical Biology and Biophysics Group, Los Alamos National Lab} 
} 
\date{}
\maketitle

\vspace{7mm}



\vspace{7mm}

\begin{abstract}
Pressure (density) and 
velocity boundary conditions inside a flow domain are studied 
for 2-D and 3-D lattice Boltzmann BGK models (LBGK) 
and a new method to specify these
conditions are proposed. These conditions are constructed 
in consistency of the wall 
boundary condition based on an idea of bounceback of non-equilibrium
distribution.  
When these  conditions 
are used together with the improved incompressible LBGK model 
\cite{zoui}, the  simulation results recover  
the analytical solution of the plane Poiseuille flow 
driven by pressure (density) difference 
with machine accuracy.
Since the half-way wall  bounceback boundary condition 
is very easy
to implement and was shown theoretically
to give second-order accuracy for 2-D Poiseuille flow with forcing,
it is used with  
pressure (density) inlet/outlet conditions proposed
in this paper  and in \cite{chen}
to study the
2-D Poiseuille flow and  the 3-D
square duct flow.  The numerical
results are approximately second-order accurate.
The magnitude of the error of the half-way wall bounceback
is comparable with  that using  
some other published boundary conditions.
Besides, the bounceback condition has a much better stability behavior
than that of other boundary conditions.

\end{abstract}


\section{Introduction}

The lattice Boltzmann equation (LBE) method has achieved great success
for simulation of transport phenomena in recent years. 
Among different LBE methods, the lattice Boltzmann BGK model is considered
more robust \cite{sdqo}. 
Some recent theoretical discussions on LBGK \cite{zou,he1}
have enhanced our understanding of the method  and the effect of boundary
conditions. 
They explains why the
velocity boundary condition for the 2-D triangular LBGK
model proposed in \cite{noble} 
generates results of machine accuracy 
for plane Poiseuille flow with forcing, and the bounceback or
equilibrium scheme generate
an first-order error to the velocity.  
Moreover, 
the bounceback scheme with the wall located half-way between a
flow node and a bounceback node (it will be called 
``half-way wall bounceback''
thereafter) is shown theoretically  
to produce results of second-order accuracy for the simple flows
considered.  

The above mentioned results with Poiseuille flows are obtained with 
external forcing to drive the flow. 
In practice, however, a flow is often  driven  by pressure difference,
and the pressure gradient in many cases cannot
be replaced by an external force in LBGK computations.
In this situation, boundary conditions usually need be implemented by
giving prescribed pressure or velocity on some ``flow boundaries'',
which are not solid walls  or interfaces of two distinct fluids.  
Instead, they are imaginary boundaries inside a flow domain 
(e.g. inlet and outlet in a pipe flow). Their existence is purely for the
convenience of study. 
The implementation of these  boundary conditions in LBGK 
is very important but it has not yet been well studied.

In lattice Boltzmann method, 
 a specification of pressure difference amounts to  a specification
of density difference.
 Early works (see, for example, \cite{daryl})
to implement pressure (density) flow boundary  condition is simply
to assign the equilibrium distribution computed with the specified
density and some velocity (maybe zero) 
to the distribution function.  
This method introduces significant errors.
Skordos \cite{sko} proposed to add a term to
the equilibrium distribution to improve it, the scheme requires the gradient of
density and velocities at boundaries.  
Inamuro {\it et al.} \cite{ina} and
Maier {\it et al.} \cite{bob}
proposed  new boundary conditions for 
lattice Boltzmann simulations. In their simulation of Poiseuille flow
with pressure (density) gradient, the pressure boundary condition
was treated  in different way compared to their wall boundary condition. 
Chen {\it et al.} \cite{chen}
also proposed a way to specify general boundary conditions
including flow  boundary conditions.

In this paper, we propose a way to specify
pressure or velocity on flow boundaries.
They are treated as one type of general boundary conditions, which
are
based on an idea of bounceback of non-equilibrium part with
modifications.
When applied to the  modified LBGK model, 
These boundary conditions produce results of machine accuracy for 
2-D Poiseuille flow with pressure (density) or velocity inlet/outlet 
conditions.
It is also noticed that although all the proposed new boundary conditions 
(\cite{chen,noble,ina,bob}) including the boundary conditions in this
paper yield improved accuracy compared to the bounceback
boundary condition,  they are difficult to implement for
general geometries,
 because there is a need to 
distinguish distribution functions according to their orientation
to the wall.
Besides, there are additional works or different treatments
at corner nodes.
On the other hand, the complete bounceback scheme does not 
distinguish  distribution functions and is very easy to implement
in a parallel way,
which was considered as one of the advantages of  LGA or LBE method. 
In this paper,
the half-way wall bounceback boundary condition with two flow boundary 
conditions are applied to the 2-D Poiseuille flow and a 3-D square duct flow
using the d2q9i and d3q15i lattice Boltzmann models respectively.
The results 
are approximately second-order accurate. The error is comparable with
that using some published boundary conditions.
Moreover, the half-way wall
bounceback boundary condition for stationary walls
is much more stable  than the boundary conditions 
in \cite{chen,noble,ina,bob} and in this paper.
Thus, we recommend to use  the half-way wall
bounceback boundary condition for stationary walls
and use the boundary conditions proposed in this paper or 
in \cite{chen,bob}  only for flow boundary conditions.

\section{Governing Equation}

 The square lattice LBGK model (d2q9) is expressed as 
(\cite{ccmm},\cite{qian2},\cite{hchen}):
\begin{equation}
f_i({\bf x}+\delta {\bf e}_i, t+\delta)-f_i({\bf x},t)=
 -\frac 1{\tau}[f_i({\bf x},t)-f_i^{(eq)}({\bf x},t)], \ \ i=0,1,...,8,
\label{eq:lbgk}
\end{equation}
where  the equation is written in physical units.
Both the time step and the lattice spacing have the value of $\delta$
in physical units.
$f_{i}({\bf x},t) $ is the density distribution
function along the direction ${\bf e}_i$ at (${\bf x}, t$).  The 
particle speed ${\bf e}_i$'s are given by 
 ${\bf e}_{i} = (\cos(\pi(i-1)/2), \sin(\pi(i-1)/2), i = 1,2,3,4$, and
${\bf e}_{i} = \sqrt{2} (\cos(\pi(i-4-\frac{1}{2})/2),
\sin(\pi(i-4-\frac{1}{2})/2), i = 5,6,7,8$.
Rest particles of type 0 with ${\bf e}_{0} = 0$ is also allowed
(see Fig.~1).
The  right hand side represents the collision term
 and $\tau $ is the single relaxation time which controls the rate
of approach to equilibrium.
The density per node, $ \rho $, and the macroscopic flow velocity,
 ${\bf u} = (u_x, u_y)$,
are defined in terms of the particle distribution function by
\begin{equation}
 \sum_{i=0}^8 f_i=\rho,  \ \ \ \
 \sum_{i=1}^8 f_i {\bf e}_i =\rho {\bf u}.
\label{eq:dens}
\end{equation}
The equilibrium distribution functions 
$f_i^{(eq)}({\bf x},t) $ 
depend only on local density and
velocity and they
can be chosen in the following form (the model d2q9 \cite{qian2}):
\begin{equation}
f_{i}^{(eq)} = t_i\rho[1+3({\bf e}_{i}\cdot
{\bf u})+\frac{9}{2}({\bf e}_{i}\cdot {\bf u})^2-\frac{3}{2}{\bf u}\cdot
{\bf u}],\;\; \; t_0=\frac49 ,\; t_i = \frac19, i =  1:4; 
\; t_i = \frac{1}{36}, i=5:8 . 
\label{eq:equil}
\end{equation}

A Chapman-Enskog procedure can be applied to Eq.~(\ref{eq:lbgk}) to derive
the macroscopic equations  of the model.  They are given by:
the continuity equation (with an error term $O(\delta^2)$ being omitted):
\begin{equation}
\frac{\partial \rho}{\partial t} + \nabla \cdot (\rho {\bf u})=0,
\label{eq:cont}
\end{equation}
and the momentum equation (with terms of  $O(\delta^2)$ and $O(\delta u^3)$
 being omitted):
\begin{equation}
\partial_{t}(\rho u_{\alpha})+\partial_{\beta}(\rho u_{\alpha} u_{\beta})
=
-\partial_{\alpha}(c_s^2 \rho)+\partial_{\beta}(2 \nu \rho S_{\alpha \beta}),
\label{eq:momen}
\end{equation}
where the Einstein  summation convention is used.
$S_{\alpha \beta}=\frac{1}{2}(\partial_{\alpha}u_{\beta}+\partial_{\beta}
u_{\alpha})$ is the strain-rate tensor. The pressure is given by
$p = c_s^2 \rho$, where $c_s$ is the speed of sound with
${\displaystyle c_s^2= \frac{1}{3}, }$
and
${\displaystyle \nu=\frac{2 \tau -1}{6} \delta}$,
with $\nu $ being the  the kinematic viscosity.
The form of the error terms and the derivation of these equations can be found
in \cite{qian3,hou1}.

For 2-D case, we will take the Poiseuille flow 
as an example to study the pressure (density) or velocity inlet/outlet
condition.
The analytical solution of Poiseuille flow
in a channel with width $2 L $ for the Navier-Stokes equation is given by:
\begin{equation}
 u_x = u_0 (1 - \frac{y^2}{L^2}) ,
\;\;\;\; u_y = 0, \;\;\;\;
 \frac{\partial p}{\partial x} = - G, \;\;\;\;
 \frac{\partial p}{\partial y} = 0, \;\;\;\;
\label{eq:pois}
\end{equation}
where the pressure gradient 
$G$ is a constant related to the centerline velocity $u_0$ by
\begin{equation}
 G = 2 \rho \nu u_0/L^2 ,
\label{eq:g}
\end{equation}
  and the flow density $\rho$ is a constant. The Reynolds number is defined
as Re = $u_0 (2L)/\nu$.

The Poiseuille flow is an exact solution of the steady-state
incompressible Navier-Stokes equations with constant density $\rho_0$:
\begin{equation}
 \nabla \cdot  {\bf u}=0.
\label{eq:a64}
\end{equation}
\begin{equation}
\partial_{\beta}( u_{\alpha} u_{\beta})
=- \partial_{\alpha} (\frac{p}{\rho_0}) + \nu \partial_{\beta \beta} u_{\alpha},
\label{eq:a65}
\end{equation}
On the other hand, the steady-state  macroscopic equations of the 
LBGK model
are different from  the incompressible Navier-Stokes equations
Eqs.~(\ref{eq:a64},\ref{eq:a65})
by terms containing  the spatial derivative of $\rho$. These discrepancies
are called  compressibility error in LBE model.
Thus,  when pressure (density) gradient drives the flow,
$u_x$ in a LBGK simulation increases  in the $x-$direction, 
the velocity profile from the simulation
is no loger a parabolic profile. 
For a fixed Mach number
($u_0$ fixed),
  as $\delta \rightarrow 0$, the velocity
of the LBGK simulation  will not converge to the velocity in
Eq.~(\ref{eq:pois}) because the compressibility error becomes
dominant. 
This makes the comparison of $u_x$ with the analytical velocity of
Poiseuille flow somehow ambiguous.

To make a more accurate study for Poiseuille flow with pressure (density)
or velocity flow boundary condition, it is better to use the improved
incompressible LBGK model proposed in \cite{zoui}.
The model (called d2q9i)
is given by Eq.~(\ref{eq:lbgk}) with the same ${\bf e}_i$ and
the following equilibrium distributions:
\begin{equation}
 f_{i}^{(eq)}= t_i  [ \rho + 3 {\bf e}_{i} \cdot {\bf v} +
  \frac{9}{2}({\bf e}_{i} \cdot {\bf v})^2 - \frac{3}{2} {\bf v}\cdot{\bf v} ],
\;  \; t_0=\frac49 ,\; t_i = \frac19, i =  1:4;
\; t_i = \frac{1}{36}, i=5:8 .
\label{eq:eqi}
\end{equation}
and 
\begin{equation}
 \sum_{i=0}^8 f_{i} = \sum_{i=0}^8 f_{i}^{(eq)} =\rho,  \ \ \ \
 \sum_{i=1}^8 f_i {\bf e}_{i} =
 \sum_{i=1}^8 f_i^{(eq)} {\bf e}_{i} ={\bf v},
\label{eq:densi}
\end{equation}
where ${\bf v}=(v_x, v_y)$ (like the momentum in the ordinary LBGK model)
is used to represent the flow velocity.
 The macroscopic equations of d2q9i in the steady-state case (apart from error
terms of $O(\delta^2)$:
\begin{equation}
 \nabla \cdot  {\bf v}=0 ,
\label{eq:conti}
\end{equation}
\begin{equation}
\partial_{\beta}( v_{\alpha} v_{\beta})
=- \partial_{\alpha}(c_s^2 \rho)+ \nu
\partial_{\beta \beta} v_{\alpha} ,
\label{eq:momeni}
\end{equation}
are exactly the steady-state incompressible 
Navier-Stokes equation with constant density $\rho_0$. In this model
d2q9i, pressure is related to the calculated density by
$c_s^2 \rho = p/\rho_0$ ($c_s^2 = 1/3$),  
 and $\nu = \frac{2\tau-1}{6} \delta$. The 
quantity $ p/\rho_0$ will be called the effective pressure. 
Although the macroscopic equations of d2q9i in the steady-state case 
has an error of $O(\delta^2)$ to the steady-state Navier-Stokes 
equation, for some special flows like the Poiseuille flow,
it is possible that this error disappears with suitable boundary 
conditions.

\section{Pressure or Velocity Flow Boundary Condition of the 2-D Square 
Lattice LBGK Model}

In this section
a new boundary condition is proposed  based on 
an idea of bounceback on non-equilibrium part 
as follows: 
take the case of a bottom node in Fig.~1,  the boundary is aligned with 
$x-$direction with 
$f_4, f_7, f_8$ pointing into the wall.
After streaming, $f_0, f_1, f_3, f_4, f_7, f_8$ are known.
Suppose that $u_x, u_y$ are specified on the wall, we 
want to use  Eqs.~(\ref{eq:dens}) to 
determine $f_2, f_5, f_6$ and $\rho$ (originated in \cite{noble}),
which can be put into  the form: 
\begin{equation}
 f_2+f_5+f_6 =
 \rho - (f_0+f_1+f_3+f_4+f_7+f_8) ,
\label{eq:rho}
\end{equation}
\begin{equation}
 f_5-f_6 = \rho u_x - (f_1-f_3-f_7+f_8) ,  \mbox{\hspace{.9in}}
\label{eq:u0}
\end{equation}
\begin{equation}
 f_2+f_5+f_6 = \rho u_y + (f_4+f_7+f_8) .\mbox{\hspace{.9in}}
\label{eq:v0}
\end{equation}
Consistency of Eqs.~(\ref{eq:rho},\ref{eq:v0}) gives
\begin{equation}
 \rho =  \frac{1}{1-u_y} [f_0+f_1+f_3+2(f_4+f_7+f_8)].
\label{eq:rhod}
\end{equation}
\par
We assume the bounceback rule is still correct for the
non-equilibrium part of the particle distribution normal to
the boundary (in this case, $f_2 - f_2^{(eq)}=f_4 - f_4^{(eq)}$).
With $f_2$ known, $f_5, f_6$ can be found, thus 
\begin{eqnarray}
f_2&=&f_4+\frac 23 \rho u_y  , \nonumber \\
f_5&=&f_7-\frac 12 (f_1-f_3)+\frac 12  \rho u_x +\frac 16  \rho u_y , \nonumber  \\
f_6&=&f_8+\frac 12 (f_1-f_3)-\frac 12\rho u_x +\frac 16 \rho u_y.
\label{eq:f256}
\end{eqnarray}
The collision step is applied to the boundary nodes also.
For non-slip boundaries, 
this boundary condition is reduced to that in \cite{bob}. 
A detailed discussion of implementation of boundary conditions
on stationary walls in 3D case was given in \cite{bob}.

\subsection{Specification of Pressure on a  Flow Boundary } 
 
Now let us turn to the pressure (density) flow boundary condition. 
In \cite{ina,bob}, the pressure (density) boundary condition was
treated in different way compared to the wall boundary condition.
Chen {\it et al.} \cite{chen} use an extrapolation scheme on an additional layer
beyond a boundary to determine the 
incoming $f_i$'s before the streaming step, their treatment of
pressure boundary condition is consistent with the wall boundary condition.
In this paper, we treat  pressure (density) flow boundary condition
the same way as the velocity wall boundary condition.
Its derivation is based on Eq.~(\ref{eq:dens}) as for velocity wall
boundary condition.
Suppose a flow boundary (take the inlet in Fig.~1 as example)
is along the $y-$direction,
and the pressure (density) is to be specified on it. Suppose that $u_y$ is also 
specified (e.g. $u_y=0$ at the inlet in a channel
flow).  
After streaming, $f_2,f_3,f_4,f_6,f_7$  are known, $\rho
= \rho_{in}, u_y=0$ are specified
at inlet. We need to  determine
$u_x$ and $ f_1, f_5, f_8$ from Eq.~(\ref{eq:dens}) as following:
\begin{equation}
 f_1+f_5+f_8 =
 \rho_{in} - (f_0+f_2+f_3+f_4+f_6+f_7) ,
\label{eq:rhoi}
\end{equation}
\begin{equation}
 f_1+f_5+f_8 =
  \rho_{in} u_x + (f_3+f_6+f_7) ,  \mbox{\hspace{.9in}}
\label{eq:u0i}
\end{equation}
\begin{equation}
 f_5-f_8 = -f_2+f_4-f_6+f_7  .\mbox{\hspace{.9in}}
\label{eq:v0i}
\end{equation}
Consistency of Eqs.~(\ref{eq:rhoi},\ref{eq:u0i}) gives
\begin{equation}
 u_x =  1 - \frac{ [f_0+f_2+f_4+2(f_3+f_6+f_7)]}{\rho_{in}} .
\label{eq:u0d}
\end{equation}
\par
We use  bounceback rule for the
non-equilibrium part of the particle distribution normal to
the inlet to find $f_1 - f_1^{(eq)}=f_3 - f_3^{(eq)}$.
With $f_1$ known,  $f_5, f_8$ are obtained by the remaining two
equations:
\begin{eqnarray}
f_1&=&f_3+\frac 23 \rho_{in} u_x , \nonumber \\
f_5&=&f_7-\frac 12 (f_2-f_4)+\frac 16  \rho_{in} u_x  ,  \nonumber \\
f_8&=&f_6+\frac 12 (f_2-f_4)+\frac 16 \rho_{in} u_x .
\label{eq:f158}
\end{eqnarray}

The corner node at inlet needs some special treatment. Take the
bottom node at inlet as an example, 
after streaming, $ f_3,f_4,f_7$  are known; $\rho$ is specified,
and $u_x=u_y = 0$. We need to determine  
$ f_1,f_2,f_5,f_6,f_8$. We use  bounceback rule for the
non-equilibrium part of the particle distribution normal to
the inlet and the boundary to find:
\begin{equation}
 f_1 = f_3 + (f_1^{(eq)}- f_3^{(eq)}) = f_3 , \;\; 
 f_2 = f_4 + (f_2^{(eq)}- f_4^{(eq)})  = f_4,  
\label{eq:f13}
\end{equation}
Using these $f_1, f_2$ in Eqs.~(\ref{eq:u0i},\ref{eq:v0i}),
we find:
\begin{equation}
 f_5 = f_7, \;\; f_6=f_8=\frac 12 [\rho_{in}-(f_0+f_1+f_2+f_3+f_4+f_5+f_7)] .
\label{eq:f5678}
\end{equation}
Similar procedure can be applied to top inlet node and outlet nodes
including outlet corner nodes.

\subsection{Specification of Velocity on a Flow Boundary }

In some calculations, velocities $u_x, u_y$ are specified at a flow boundary 
(take the inlet in Fig.~1 as example). 
In the flow region of the inlet or outlet,
this is actually equivalent to a velocity wall boundary 
condition and can be handled in the same way as given at the beginning
of the section.
The effect of specifying  velocity at inlet is similar to specifying 
pressure (density) at inlet. Density difference in the flow
can be automatically generated by the velocity inlet condition.   
 
At the inlet bottom (non-slip boundary), special treatment is needed. 
After streaming,
$f_1, f_2, f_5, f_6, f_8$ need to be determined. Using bounceback
on normal distributions gives:
\[ f_1 = f_3, \;\; f_2 = f_4 .   \]
Expressions of $x, y$ momenta give:
\begin{eqnarray}
f_5-f_6+f_8 = -(f_1-f_3-f_7)= f_7 , \nonumber \\
f_5+f_6-f_8 = -(f_2-f_4-f_7)= f_7 ,
\label{eq:veli}
\end{eqnarray}
or
\begin{eqnarray}
f_5&=&f_7 ,  \nonumber \\
f_6&=&f_8 = \frac 12 [\rho - (f_0+f_1+f_2+f_3+f_4+f_5+f_7) ] ,    
\label{eq:veli1}
\end{eqnarray}
but there is no more equation available to determine $\rho$. The situation
is similar to a corner wall node (the intersection of two perpendicular
walls).
In this situation, since $\rho$ is expected to be constant at the
inlet, $\rho$ at the inlet bottom node can be taken as the
$\rho$ of its neighboring  flow node, thus the velocity inlet condition
is specified. 
It is noted that this treatment is only for a special study to produce
the analytical solution with model d2q9i. 
In practice, however,
the half-way wall bounceback is recommeded as boundary conditions
for stationary walls and there is no need for special treatment at
corner nodes. This will be discussed in section 4. 

From the discussion given above, we can unify boundary conditions (on a wall
boundary or in a flow boundary) in 2-D 
simulation on a straight boundary as:
\begin{itemize}
\item  Given $u_x, u_y$, find $\rho$ and unknown $f_i$'s.
\item  Given $\rho$ and the velocity along the boundary, find the velocity
normal to the boundary and unknown $f_i$'s.
\end{itemize}

The above discussion is for  flat flow boundaries  aligned with
a plane  spanned by  two particle  velocities of type I.
It is not the purpose of this paper to derive a flow
boundary  condition in a general geometry.
If a flow boundary has  a complicated geometry or if
it is not aligned with lattice directions, schemes based on extrapolations 
like the ones in \cite{chen,bob} can be used. 
However, as pointed out in \cite{bob}, on a convex edge or 
at a convex corner, there are too few unknown $f_i$'s, if
the available $f_i$'s are used, then any choice of
the unknown $f_i$'s may not 
 make the velocity correct.  There have not been analytical
or numerical studies on the order of accuracy for these situations.
Thus the order of 
accuracy is not clear for these cases. 
Hence, the equilibrium scheme  can be considered
as well.  The equilibrium scheme at boundaries gives second-order 
accuracy if $\tau = 1$ but only first-order when $\tau \neq  1$
\cite{sko, noble2}.

\subsection{Boundary Conditions for the Modified Incompressible Model d2q9i}

The velocity wall boundary condition and flow boundary conditions for d2q9i 
are similar to
that of d2q9. The derivation is based on
 equations $\sum_{i=0}^8 f_i = \rho$ and
$\sum_{i=1}^8 {\bf e}_i f_i = {\bf v}$ and hence some modifications are 
needed as follows:
\begin{itemize}
\item In wall boundary condition, Eq.~(\ref{eq:rhod}) is replaced by
\begin{equation}
 \rho = v_y + [f_0+f_1+f_3+2(f_4+f_7+f_8)].
\label{eq:rhodi}
\end{equation}
\par
and in Eq.~(\ref{eq:f256}), $\rho u_x, \rho u_y$ are replaced by
$v_x, v_y$ respectively. 
\item In pressure flow boundary condition, Eq.~(\ref{eq:u0d}) is replaced by
\begin{equation}
 v_x =  \rho_{in} - [f_0+f_2+f_4+2(f_3+f_6+f_7)] ,
\label{eq:u0di}
\end{equation}
and in Eq.~(\ref{eq:f158}), $\rho_{in} u_x $ is replaced by
$v_x $.

\end{itemize}

\subsection{Numerical Results of Model d2q9i }

We report and discuss  the numerical results for Poiseuille flow with
our wall and flow  boundary conditions. The simulation
is performed on the model d2q9i.
The main result in the simulation is the achievement of machine
accuracy. 
The width of the channel is assumed to be $2 L = 2$.
We use $nx, ny$ lattice nodes on the $x-$ and $y-$directions, thus,
$\delta = 2/(ny-1)$. 
The initial condition is to assign $f_i = f_i^{(eq)}$ computed using
a constant density $\rho_0$, and zero velocities.
The steady-state is reached if
\begin{equation}
  \frac{ \sum_{i} \sum_{j} | v_x(i,j, t+\delta) - v_x(i,j,t)|
 + | v_y(i,j, t+\delta) - v_y(i,j,t)|}
 { \sum_{i} \sum_{j} |  v_x(i,j,t)|
 + |  v_y(i,j,t)|}  \leq  \delta \cdot Tol .  
\label{eq:stea}
\end{equation}
$Tol$ is a tolerance
set to $10^{-12}$ in this section.

We also define a maximum relative  error of velocity $(v_x, v_y)$ as
in \cite{noble2}:
\begin{equation}
 err_m \equiv  \max  \frac{ \sqrt{ ( u_x^t - v_x)^2
 + ( u_y^t - v_y)^2}}
 { u_0} ,
\label{eq:err1}
\end{equation}
where $u_x^t, u_y^t$ is the analytical velocity, and $u_0$ is
the peak velocity, the maximum is throughout the flow.

For model d2q9i, we carried out
simulations with a variety of Re,  $nx, ny, u_0$ 
 using the pressure or velocity flow
boundary condition. All simulations in this paper use double-precision.
The range of Re is from 0.0001 to 30.0;
the range of $\tau $ is from 0.56 to 20.0 and 
the range of $u_0 $ is from 0.001 to 0.4;
the largest density difference simulated (not the limit)
is $\rho_{in} = 5.6, \ \rho_{out}=4.4$ with $nx=5, ny=3$ corresponding to
an effective pressure gradient of $G'=0.1$, where $G'$ is defined as
$G'= -\frac{1}{\rho_0}\frac{dp}{dx} $.
The magnitude of average density
$\rho_0$ is 5, but it is irrelevant for the simulation \cite{zoui}.

For all cases where the simulation
is stable, the steady-state  velocity and density show:
\begin{itemize}
\item The velocity field $v_x$ is uniform in
the $x$-direction, it is accurate up to machine accuracy
compared to the analytical solution in Eq.~(\ref{eq:pois}), 
$v_y$ is very small with maximum of $|v_y|$ in the whole region being 
in the order
of $10^{-13}$. For example,  for 
$nx = 5, ny = 3, u_0= 0.1, \tau=0.56, $ Re =10,  the maximum
relative error
of velocity  is $0.1816 \cdot 10^{-11}$,
while  the maximum relative error of density is $0.3553 \cdot 10^{-15}$,
and the maximum magnitude of $v_y$ is $0.5551 \cdot 10^{-15}$.
The results  for other cases are similar to this example. 
\item The density is uniform in the cross channel direction, and linear in
the flow direction. 
the computed value and the analytical value of the 
density gradient differ only at the 14th digit.
\end{itemize}

It is also noticed that with pressure (density) gradient to drive
Poiseuille flow, the maximum Reynolds
number which makes the simulation stable is far less than that with 
external forcing,  which is also reported in \cite{chen}.  

Similar results with machine accuracy
are obtained by specifying the analytical velocity profile
given in Eq.~(\ref{eq:pois})
at inlet and pressure (density) at outlet by using the flow boundary
conditions in this paper.  In the case, there is a uniform
pressure (density) difference in the region. The value of the density difference
depends on $u_0$ and the outlet density. 

The accurate results in the model d2q9i give  us confidence about 
the flow boundary conditions proposed.

\section{On Half-way Wall Bounceback  Boundary Condition}

\subsection{Reconsider Bounceback}

The bounceback rule with wall placed at the bounceback nodes 
gives an first-order error to velocity, both at the boundary and
throughout  the flow. This has been showed analytically \cite{he1}
for some simple flows and computationally \cite{hou1} for 
2D cavity flows. The same can be said for the flows studied in
this paper based on numerical results. 
There have been efforts to replace the bounceback boundary condition
with  new boundary conditions  
for LBE simulation.
 Various boundary conditions  \cite{chen,noble,sko,ina,bob,noble2} including
the one in this paper are proposed.
 While these boundary conditions indeed improve the accuracy for some
simple flows over the bounceback scheme, extra works  are still 
needed to apply them to a 
general situation.
First, for example, at a concave edge \cite{bob}, the boundary conditions 
in \cite{noble,sko,ina,bob,noble2} and in this  paper
do not give a natural way to determine the density, and since 
there are more unknowns  at a node than in a plane wall 
node, some additional assumptions are needed to apply the boundary
condition. 
Second, all the boundary conditions in
 \cite{chen,noble,sko,ina,bob,noble2}
 including
the one in this paper require different treatments 
on $f_i$'s depending on  their orientation to the wall,
the implementation of these boundary conditions is formidable 
for  complicated  geometry like that in a porous media or for
a situation where a wall is not aligned with any of the $f_i$'s 
direction.
On the other hand, the ``complete''  bounceback scheme which assigns 
each $f_i$ the value of the $f_j$ of its opposite direction
with no relaxation on the bounceback nodes
is very easy and convenient to  apply, the treatment is independent on 
the direction of $f_i$'s, which is one of the major advantages of
LGA or LBE method.
Several theoretical
studies have shown that if the 
 wall is placed at the half way
between the bounceback row and the first flow row
(``half-way wall bounceback''), the scheme 
gives a second-order accuracy  \cite{zou,he1,corn,ginzbourg94}
for some simple flows including an inclined channel flow and a
plane stagnation flow.
For example, if the d2q9 model with
the half-way wall bounceback is used to simulate 
the 2-D Poiseuille flow with forcing, the 
error of the velocity (it is the same for any node) is given by \cite{he1}
\begin{equation}
 u_j^t - u_j = - \frac{u_0 [4 \tau (4\tau-5)+3]}{3} \delta^2 
\label{eq:errp}
\end{equation}
where $u_j^t, u_j$ is the analytical and computed $x$-velocity 
respectively, $u_0$ is the center velocity.
For a fixed $\tau$, the error  is second-order 
 in the lattice spacing $\delta$.  Of course, large value of  $\tau$ 
will give large errors, which are reported in some papers
\cite{noble,ina}, giving an impressed vision how bad the bounceback
boundary condition was. 
However, for practical purposes, 
 there is no need to take large value of
$\tau$ in a simulation. Besides, the Chapman-Enskog procedure 
does not work right when $\tau$ is large.
For $\tau$ between 0.5 and 1.25, the magnitude of the error given 
in (\ref{eq:errp}) is less than or equal to 1.1 $u_0 \delta^2$.
Thus, 
it is worthwhile to consider this boundary condition in more general
situations especially in 3-D flows and some results are reported in
section 4.2, 4.3  and section 5.

\subsection{Results of Model d2q9i}

To reduce the effect of compressibility error, the model d2q9i is
used.  Simulations with model d2q9 were also carried out, the results
have very similar behavior for order of convergence but the 
magnitude of error in d2q9 is greater.
We use d2q9i to simulate  Poiseuille flow with pressure or velocity flow 
boundary condition.  For the half-way wall bounceback, 
if there are $ny$ nodes on the $y-$direction, then the first and last
nodes are the bounceback nodes with the wall 
being located half-way between the bounceback node and the first flow node, 
 there are $ly=ny-2$ lattice steps across the
channel and the lattice spacing is $\delta = 2/(ny-2) = 2/ly$
(in the case of the boundary condition in \cite{chen}
$\delta = 2/(ny-1) = 2/ly$).
The length of the channel is set to twice 
as the width.  
 At the inlet/outlet,
the bounceback is also used at the nodes on the
bounceback rows,  thus, there is
no additional treatment for the corner nodes at the inlet/outlet.

Two inlet/outlet (I/O) conditions with the half-way wall bounceback were tested:
\begin{enumerate}
 \item I/O No. 1:  the flow boundary condition given 
in this paper.  
 \item I/O No. 2:  the flow boundary condition proposed in
\cite{chen}.  It assumes an additional layer of nodes beyond the boundary
flow nodes and uses  an extrapolation  formula to derive the incoming
$f_i$'s of the additional layer before streaming.
\end{enumerate}
The I/O and boundary condition in \cite{chen} were also used to
compare result with the the half-way wall bounceback.

For the study of accuracy, we fix $\tau$ and the Reynolds number
(which affects stability) and halve $\delta$ 
each time and calculated the error in the result.
Simulations with different Re and $\tau \leq 1.3$  are performed.
Three examples with pressure specified at inlet and outlet
are reported on Table I,
The example uses three sets of parameters: (1) $\tau=0.6$, Re = 10 ,
(2) $\tau=0.8, $ Re = 10, and
(3) $\tau=1.1, $ Re = 1
 (Re is restricted so that
$u_0$ for the smallest number of lattice nodes is still small), and 
$\rho_0 = 5$.  The quantity $Tol$ in Eq.~(\ref{eq:stea}) is set 
to $10^{-8}$.
As $\delta$ halves, $u_0$ also halves, so does the Mach number,
the same way as in the study of duct flow in \cite{bob}.
As $\delta$ changes, $\nu$  and
the effective pressure gradient $G'$ 
and then pressure (density) at inlet/outlet also changes.
$lx, ly$ are used to  represent the number of lattice steps in
$x-$ and $y-$directions, we use $lx = 8,16,32,64,128, 
ly = 4,8,16,32,64$ respectively to do the
simulation. 

The convergence result is summerized in Table~I. 
The ratio of two consecutive  maximum relative errors is also shown.  The order of
convergence from a least-square fitting (a linear least-squares fitting
to logarithms of error and $\delta$)  is shown in the last column.

For the cases of I/O Nos. 1, 2 with 
the half-way wall bounceback,  
the maximum relative velocity errors are very close,
the ratio is approximately equal to 4, indicating a second-order accuracy.
The magnitude of errors in  the
half-way wall bounceback is close to 
(sometimes smaller than) 
the result using I/O and boundary condition in \cite{chen}. 
For the cases of  the half-way wall bounceback, it is also observed that
\begin{itemize}
\item Velocity $v_x$ is generally uniform in the $x-$direction.
\item  $v_y$ is small compared to $u_0$,
with maximum of $|v_y|$ being less than 
$ 0.011 u_0$ in all cases. As $ly$ increases, the  maximum of 
$|v_y|/u_0$ decreases.
I/O No. 2 gives much smaller $\max |v_y|$ than I/O  No. 1. 
\item The density is approximately
uniform in the cross channel direction, and linear in
the flow direction. 
The maximum relative density error is much less than the maximum relative velocity error and
it decreases much faster than the maximum relative velocity error 
as $ly$ increases. 
The density gradient $(\rho(i+1,j) - \rho(i,j))/\delta$
is approximately equal to the
analytical value.  
 No. 2 gives almost the exact density distribution. 
\end{itemize}

Of course, the achievement of second-order accuracy for
Poiseuille flow does not necessarily 
mean a second-order accuracy for any flow, in next section,
a 3-D duct flow will be studied.

\subsection{Stability Issue}

Another important issue is the stability related to boundary conditions.
It is found that the combination of
bounceback  without collision on stationary
walls with equilibrium distribution 
at flow boundaries  gives the best behavior
on stability. 
Once any boundary condition or flow boundary  condition in 
any of the schemes in \cite{chen,noble,ina,bob,noble2} or
in this paper is used, the maximum Re number is reduced dramatically.
For example, in the simulation of Poiseuille flow
with bounceback at the wall 
and equilibrium scheme with velocity inlet and density outlet 
conditions  (in the case, if density are prescribed at both inlet and 
outlet, the velocity is significantly smaller than the intended value
for high Re flows) 
and with  $lx=16, ly=8$, $u_0 = 0.1$, the maximum Re
is 500.  The maximum Re reduces to 63, 56 respectively for
density inlet/outlet condition  No. 1, No. 2 with bounceback boundary condition
on the walls. 
 The maximum Re further reduces to 42, 12 respectively for
density inlet/outlet condition and boundary conditions in this paper and
in \cite{chen}. 
It is also noted that when the parameter are close to the region of instability,
the simulation may have unusual large errors.  On this account, the 
half-way wall
bounceback is safer.

\section{Flow Boundary Conditions and
Results  for the 3-D 15-velocity 
LBGK Model}

Since 3-D model is needed in practical problems, this section will
 discuss
the pressure or velocity flow boundary condition for the 3-D 
15-velocity LBGK model d3q15 and an incompressible model
d3q15i similar to d2q9i
and present some simulation results. 
The model d3q15 is based on the LBGK equation Eq.~(\ref{eq:lbgk})
with $i = 0, 1, \cdots, 14$, where ${\bf e}_i, i = 0, 1, \cdots, 14$ are the 
column vectors of the following matrix:
\[ E =  \left[ \begin{array}{rrrrrrrrrrrrrrr}
  0 &1 &-1 &0 &0 &0 &0 &1 &-1 &1 &-1 &1 &-1 &1 &-1       \\
  0 &0 &0 &1 &-1 &0 &0 &1 &-1 &1 &-1 &-1 &1 &-1 &1       \\
 0 &0 &0 &0 & 0 &1 &-1 &1 &-1 &-1 &1 &1 &-1 &-1 &1       
         \end{array}    \right]               \]
and ${\bf e}_i, i = 1, \cdots, 6$  are classified as type I, 
${\bf e}_i, i = 7, \cdots, 14$  are classified as type II.
The density per node, $ \rho $, and the macroscopic flow velocity,
 ${\bf u} = (u_x, u_y, u_z)$,
are defined in terms of the particle distribution function by
\begin{equation}
 \sum_{i=0}^{14} f_i=\rho,  \ \ \ \
 \sum_{i=1}^{14} f_i {\bf e}_i =\rho {\bf u}.
\label{eq:3ddens}
\end{equation}
The equilibrium can be chosen as:
\begin{eqnarray}
f_{0}^{(eq)}&=&\frac{1}{8}\rho- \frac 13 \rho {\bf u}\cdot{\bf u} ,
\mbox{\hspace{1.65 in}}
\nonumber\\
f_{i}^{(eq)}&=&\frac{1}{8}\rho +\frac 13 \rho{\bf e}_{i}\cdot
{\bf u}+ \frac{1}{2}\rho({\bf e}_{i}\cdot {\bf u})^2-\frac{1}{6}\rho
{\bf u}\cdot {\bf u} ,\;\; i \in  I \nonumber \\
f_{i}^{(eq)}&=&\frac{1}{64}\rho +\frac{1}{24} \rho{\bf e}_{i}\cdot
{\bf u}+ \frac{1}{16}\rho({\bf e}_{i}\cdot {\bf u})^2-\frac{1}{48}\rho
{\bf u}\cdot {\bf u}  .\;\; i \in  II 
\label{eq:3dequil}
\end{eqnarray}
The model d3q15i is constructed from these formulas in a
similar  way as in d2q9i.

The flow to be studied in the 3-D case is the square duct flow
with $x-$ direction  being the flow direction. 
It is 
clear to give a projection of the velocities  in the $xz$ plane
as shown in Fig.~1. 
The macroscopic equations of the model is the same as 
Eqs.~(\ref{eq:cont},\ref{eq:momen}) with $c_s^2 = 3/8$, and 
$\nu = (2 \tau -1)\delta /6$.

The pressure flow 
boundary condition proposed in  3.1 has the following version 
for the model d3q15:
take the case of an  inlet node  as shown in Fig.~1, the inlet
is on $yz$ plane.
After streaming, $f_i, (i = 0,2,3,4,5,6,8,10,12,14)$ are known,
Suppose that $\rho_{in}, u_y=u_z=0$ are specified on the inlet, we need to
determine $f_i, i = 1,7,9,11,13$
 and $u_x$ from Eqs.~(\ref{eq:3ddens}).
Similar to the derivation in d2q9, $u_x$ is determined by a consistency
condition as:
\begin{equation}
 \rho_{in} u_x = \rho_{in} - [f_0+f_3+f_4+f_5+f_6+2(f_2+f_8+f_{10}+f_{12}+
f_{14})  ] ,
\label{eq:3din1}
\end{equation}
The expression of $x-$ momentum gives:
\begin{equation}
 f_1+f_7+f_{9}+f_{11}+f_{13} = 
 \rho_{in} u_x + (f_2+f_8+f_{10}+f_{12}+f_{14})  ,
\label{eq:3dux}
\end{equation}
If we use  bounceback rule for the
non-equilibrium part of the particle distribution 
$ f_i, (i=1,7,9,11,13)$ to set
\begin{equation}
f_i = f_{i+1} + (f_i^{(eq)} - f_{i+1}^{(eq)}) , \; 
\label{eq:bbn}
\end{equation}
then Eq.~(\ref{eq:3dux}) is satisfied, and all $f_i$ are defined.
In order to get the correct  
$y-, z-$momenta,  we further fix this $f_1$
(bounceback of non-equilibrium $f_i$ in the normal direction)
and modify 
$f_7, f_{9}, f_{11}, f_{13}$  as in \cite{bob}:
\begin{equation}
  f_i \leftarrow f_i + \frac 14 e_{iy}\delta_y +\frac 14 e_{iz} 
\delta_z . \;\; i = 7, 9, 11, 13   
\label{eq:bbm}
\end{equation}
This modification  leaves $x-$momentum unchanged   but adds 
$\delta_y, \delta_z$ to the $y-, z-$momenta respectively. A suitable
choice of $\delta_y$ and $\delta_z$ then gives the correct
$y-, z-$momenta. Finally, we find:
\begin{eqnarray}
 f_1&=&f_2 +  \frac 23 \rho_{in} u_x ,  \nonumber \\
 f_i&=&f_{i+1} +  \frac{1}{12} \rho_{in} u_x - \frac 14 [e_{iy}(f_3-f_4) +
 e_{iz}(f_5-f_6) ] , \;\; i = 7,9,11,13
\label{eq:3din2}
\end{eqnarray}
There is no special treatment at the wall of the inlet/outlet if
bounceback is used there. Modification of the flow boundary condition
for d3q15i is similar to d2q9i.

The velocity flow boundary  condition can be derived similarly.

Simulations on 3-D square duct flow are performed on d3q15i and
on d3q15 
using the pressure flow boundary condition. Only the result of
d3q15i is reported, the result of d3q15 is similar but the errors
are greater.
The analytical solution of a flow in an infinitely long rectangular
duct $-a \leq y \leq a, \ -b \leq z \leq b, $ with $x$ being the 
flow direction 
is given by \cite{white}
\begin{equation}
  u_x(y,z) = \frac{16a^2}{\mu \pi^3} (-\frac{dp}{dx})
\sum_{i=1,3,5, \cdots}^{\infty}
(-1)^{(i-1)/2} [ 1 - \frac{\cosh(i\pi z/2a)}{\cosh(i\pi b/2a)} ]
  \frac{\cos(i\pi y/2a)}{i^3} . \ \ 
\label{eq:duct}
\end{equation}
We use $a=b=2$ in the simulations.
Results of the following boundary conditions are reported:
(1) I/O No. 1 (the I/O condition discussed above)
with  the half-way  wall bounceback at walls; \\
(2) I/O No. 2 (the I/O condition in \cite{chen}) with the 
half-way wall bounceback 
at walls; \\
(3) I/O and boundary conditions in \cite{chen}.   \\
(4) I/O and boundary conditions in \cite{bob}. \\
(5) I/O No. 2 with the original bounceback, 
which  is the bounceback at wall nodes
without collision (only for the middle example). \\
The I/O No. 1 with solid boundary condition in [10] is also tried,
the result on velocity is very close to that in case (4) and is not reported.
Again, we fix $\tau$ and the Reynolds number,
 halve $\delta$ each time.
Simulations with different Re and $\tau \leq 1.3$  are performed.
Three examples are reported on Table II:
(1) $\tau=0.6$, Re = 10, 
(2) $\tau=0.8$, Re = 5, 
(3) $\tau=1.1$, Re = 0.2, 
and all use $\rho_0 = 5$.  
The quantity $Tol$ in the 3-D version of
Eq.~(\ref{eq:stea})  is set
as $10^{-8}$.
Define $lx, ly, lz$ to represent the number of lattice steps in
$x-$ and $y-,z-$directions respectively, we use $lx = 8,16,32,64,
ly=lz= 4,8,16,32$ respectively to do the
simulation.
The maximum relative error is defined similar to  Eq.~(\ref{eq:err1}).
From Table II, we can see that the results of 
the half-way wall bounceback 
give an accuracy close to second-order, so  do the boundary
conditions in \cite{chen,bob}.
 Besides, the errors
with the half-way wall bounceback are comparable with
that using the I/O and boundary conditions in \cite{chen,bob}.
On the other hand, the original bounceback introduces  an error of 
first-order throughout the flow (L1 error has a similar 
first-order behavior).  
Thus, the half-way wall bounceback has an essentially different
behavior than the original bounceback.  
The calculated density  has smaller errors than that of the velocity,
and the density is less uniform across a section in the duct than
in 2-D case.

It is noted that the orders of convergence in the case of  3-D
duct flow with all boundary conditions considered  
are not as good as in 2-D Poiseuille flows.
For example, the ratios of errors at the highest resolution are not 
very close to 4 in some cases (the order of convergence may be close to 2 in the
cases because of an irregular large error ratio at a coarser
resolution).
It looks that in 3-D duct flow, the four edges pose additional  difficulties 
to boundary conditions. Even with forcing, the density is not 
uniform in a cross section for the half-way wall bounceback or for 
boundary conditions in \cite{chen,bob}. 
Nevertheless, the order of convergence for 3-D duct flow 
is still close to 2. The half-way wall bounceback has a weaker convergence 
when $\tau >1$  while the boundary condition in
\cite{chen} performs better as $\tau > 1$ but worse as $\tau < 1$.
 On the consideration of
simplicity in implementation and superior stability behavior
of the half-way wall bounceback,
we think that it deserves a serious consideration in a LBGK simulation.

We would like to point out that
a second-order accuracy of the half-way wall bounceback in the flows
considered does not imply a second-order accuracy for
any flows.  The statement  can be applied to  other boundary conditions
as well.  One is encouraged to do some tests on a simplified flow
of the type of flows to be simulated.

\section{Discussions}

 The major results in this paper are: first, flow boundary conditions
can be treated in a similar way as  wall boundary conditions, and 
a new way to specify flow boundary conditions 
based on bounceback of non-equilibrium part  is proposed.  For the test
problem of Poiseuille flow 
with  pressure or velocity inlet/outlet conditions,
the new method recovers the analytic solution within machine accuracy.
Second, we reconsider the the half-way wall bounceback  boundary condition for
stationary walls.  It is very easy to implement.
If being applied with flow boundary condition in this
paper or  in \cite{chen},
it is  
approximately  of second-order accuracy for the 2-D and 
3-D channel flows
with $\tau $ being less than or close to one. The magnitude of the error is 
comparable with that using some published
boundary conditions.
Hence, the the half-way wall bounceback is recommended for stationary walls.
For flow boundary conditions in simulations of small to
moderate Re numbers, 
one may consider the schemes in this paper or in
\cite{chen}.
For the cases considered, these flow boundary conditions give
very close results in velocity. The  scheme in \cite{chen}
gives a  better density distribution but a weaker  stability behavior.
One can also use the equilibrium distribution scheme  
if $\tau=1$ can be used.
However, for simulations of large Re flows, one may have to use the
equilibrium flow boundary condition  
with  velocity inlet condition and with $\tau $ close to 0.5. 
In that situation, the accuracy is only first-order.

\section{Acknowledgments}

Discussions with R. Maier, R. Bernard  are appreciated. 
Q.Z. would like to thank the Associated Western Universities  Inc.
for providing a fellowship and to thank G. Doolen and S. Chen 
for helping to arrange his visit to
 the Los Alamos National Lab. 
Some computations are performed on the Convex Exempler SPP-1000
of Kansas State University. Q.Z. would like to thank NSF 
 grant number DMR-9413513 which provided
fund for the acquisition of the machine.

\vfill\eject


\newpage

 Table I. maximum relative errors for the 2-D Poiseuille flow 
with pressure being specified at inlet and outlet
for three cases: (1) Re=10, $\tau=0.6$,
 (2) Re=10, $\tau=0.8$,
 (3) Re=1, $\tau=1.1$.
The results of  
two pressure flow boundary conditions I/O Nos. 1,2 
with the half-way wall bounceback (HWWBB) and the result
of the I/O and boundary conditions in [2] are given. 
In each box, the upper figure is the error,
and the lower figure is the ratio of two consecutive errors.
The last column shows the order of convergence using the
least-squares fitting.  The symbol (-2) represents $10^{-2}$. \\

\begin{tabular}{|l|c|l|l|l|l|l|l|}  \hline
  &lx              &8 &16   &32  &64  &128 &order    \\  
  &ly              &4 &8 &16   &32  &64    &  \\ \hline 
  &$u_0$     &0.8333(-1)  &0.4167(-1) &0.2084(-1) &0.1042(-1)
  &0.5208(-2) & \\  \cline{2-8}
  &I/O No. 1 &0.6031(-1)&0.1500(-1)&0.3729(-2) &0.9297(-2) &0.2324(-3) &2.005
\\
Re = 10 &HWWBB &4.021  & 4.023  &4.011  &4.000   &   &  \\ \cline{2-8}
$\tau=0.6$ &I/O No. 2  &0.5917(-1)&0.1479(-1) &0.3699(-2) &0.9265(-2) &0.2352(-3
) &1.995  \\
       &HWWBB &4.001  & 3.998  &3.992  &3.939   &   &     \\ \cline{2-8}
  &I/O, B.C.  & \multicolumn{6}{c|}{unstable}      \\
  &in [2] &  \multicolumn{6}{c|}{}                 \\ \hline\hline
  &$u_0$     &0.2500  &0.1250 &0.6250(-1) &0.3125(-1)
  &0.1563(-1) & \\  \cline{2-8}
  &I/O No. 1 &0.3276(-1)&0.8319(-2)&0.2054(-2) &0.5111(-3) &0.1276(-3) &2.003   \\ 
Re = 10 &HWWBB &3.938  & 4.050  &4.019  &4.005   &   &  \\ \cline{2-8}
$\tau=0.8$ &I/O No. 2  &0.3250(-1)&0.8125(-1) &0.2032(-2) &0.5085(-3) &0.1283(-3) &1.997  \\ 
       &HWWBB &4.000  & 3.999  &3.996  &3.963   &   &     \\ \cline{2-8}
  &I/O, B.C.  &0.1000 &0.2500(-1)&0.6250(-2) &0.1563(-2) &0.3920(-3) &1.999  \\ 
  &in [2] &4.000  & 4.000  &3.999  &3.987   &   &  \\ \hline\hline
  &$u_0$     &0.5000(-1) &0.2500(-1) &0.1250(-1) &0.6250(-2)
  &0.3125(-2) & \\  \cline{2-8}
  &I/O No. 1 &0.5550(-1)&0.1441(-1)&0.3617(-2) &0.9021(-3) &0.2249(-3) &1.989   \\ 
Re = 1 &HWWBB &4.011  & 4.010  &3.984  &3.851   &   &  \\ \cline{2-8}
$\tau=1.1$ &I/O No. 2  &0.5750(-1)&0.1437(-1)&0.3594(-2) &0.8984(-3) &0.2246(-3) &2.000  \\ 
       &HWWBB &4.001  & 3.998  &4.000  &4.000   &   &     \\ \cline{2-8}
 &I/O, B.C.  &0.5000(-1) &0.1250(-1)&0.3125(-2) &0.7812(-3) &0.1953(-3) &2.000  \\ 
  &in [2] &4.000  & 4.000  &4.000  &4.000   &   &  \\ \hline\hline
\end{tabular}

\vspace{10mm}

\newpage

 Table II. maximum relative errors for the 3-D square duct flow
with pressure being specified at inlet and outlet
for three cases: (1) Re=10, $\tau=0.6$,
 (2) Re=5, $\tau=0.8$,
 (3) Re=0.2, $\tau=1.1$.
The results of
two pressure flow boundary conditions I/O Nos. 1,2
with the half-way wall bounceback (HWWBB), the results
of the I/O and boundary conditions in [2,10] are also given.
 The result of I/O in [2]
with original bounceback  is given for the second case.
In each box, the upper figure is the error,
and the lower figure is the ratio of two consecutive errors.
The last column shows the order of convergence using the
least-squares fitting.   \\

\begin{tabular}{|l|c|l|l|l|l|l|}  \hline
  &lx              &8 &16   &32  &64   &order    \\
  &ly,lz           &4 &8 &16   &32      &  \\ \hline
  &$u_0$    &0.8333(-1)&0.4167(-1)&0.2083(-1) &0.1042(-1) & \\  \cline{2-7}
  &I/O No. 1 &0.4028&0.1054&0.2742(-1) &0.7289(-2) &1.931   \\
Re = 10 &HWWBB &3.822  & 3.844  &3.762     &   &  \\ \cline{2-7}
$\tau=0.6$ &I/O No. 2 & \multicolumn{5}{c|}{unstable}   \\
       &HWWBB  &  \multicolumn{5}{c|}{}    \\ \cline{2-7}
  &I/O, B.C.  & \multicolumn{5}{c|}{unstable}   \\
  &in [2] & \multicolumn{5}{c|}{}    \\ \cline{2-7}
  &I/O, B.C.  &0.2210 &0.5565(-1)&0.1293(-1) &0.3132(-2)  &2.053  \\
  &in [10] &3.971  & 4.304  &4.128     &   &  \\ \hline\hline
  &$u_0$       &0.1250 &0.6250(-1) &0.3125(-1) &0.1563(-1) & \\  \cline{2-7}
  &I/O No. 1 &0.1382&0.3980(-1)&0.9805(-2) &0.2388(-2) &1.959   \\ 
Re = 5 &HWWBB &3.472  & 4.059  &4.106     &   &  \\ \cline{2-7}
$\tau=0.8$ &I/O No. 2 &0.1371&0.3659(-1)&0.9243(-2) &0.2310(-2) &1.966   \\ 
       &HWWBB  & 3.747  &3.959  &4.001   &    &    \\ \cline{2-7}
  &I/O, B.C.  &0.3397 &0.9563(-1)&0.2117(-1) &0.5741(-2)  &1.984  \\
  &in [2] &3.552  & 4.517  &3.688     &   &  \\ \cline{2-7}
  &I/O, B.C.  &0.8567(-1) &0.1543(-1)&0.4502(-1) &0.1278(-2)  &1.998  \\
  &in [10] &5.552  & 3.427  &3.523     &   &  \\ \cline{2-7}
  &I/O No. 2 &0.6539 &0.3345&0.1536 &0.7935(-1)  &1.025  \\
  &bounceback   &1.955  & 2.178  &1.936     &   &  \\ \hline\hline
  &$u_0$  &0.1000(-1)  &0.5000(-2) &0.2500(-2) &0.1250(-2) & \\  \cline{2-7}
  &I/O No. 1 &0.2091&0.6537(-1)&0.1817(-1) &0.4807(-2) &1.818   \\
Re = 0.2 &HWWBB &3.199  & 3.598  &3.780     &   &  \\ \cline{2-7}
$\tau=1.1$ &I/O No. 2 &0.2114&0.6448(-1)&0.1787(-1) &0.4737(-2) &1.829   \\
       &HWWBB  & 3.279  &3.608  &3.772   &    &    \\ \cline{2-7}
  &I/O, B.C.  &0.3109 &0.7740(-1)&0.1966(-1) &0.4904(-2)  &1.994 \\
  &in [2] &4.017  & 3.937  &4.009     &   &  \\ \cline{2-7}
  &I/O, B.C.  &0.2070 &0.4973(-1)&0.1277(-1) &0.3341(-2)  &1.982  \\
  &in [10] &4.162  & 3.894  &3.822     &   &  \\ \hline\hline
\end{tabular}

\newpage

\section{Figure Caption}

Fig.~1, Schematic plot of velocity directions of the 2-D (d2q9) model
and projection of 3-D (d3q15) model in a channel.
In the 3-D model, The $y-$axis is pointing into the paper, so are
velocity directions 3,7,9,12,14 (they are in parentheses if shown), while the
velocity directions 4,8,10,11,13 are pointing out. Velocity directions
3,4 have
a projection at the center and are not shown in the figure.

\vspace{1in}

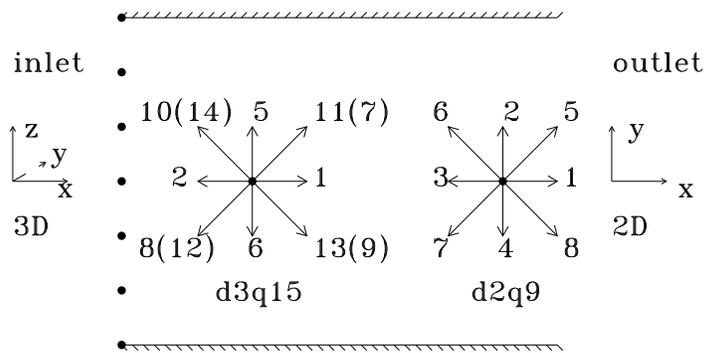
\end{document}